\documentstyle[aps,prd,preprint,fleqn]{revtex}
\begin{document}
\draft
\title{Global effects in quaternionic quantum field theory}
\author{S.\ P.\ Brumby and G.\ C.\ Joshi}
\address{Research Centre for High Energy Physics,\\
School of Physics, University of Melbourne,\\
Parkville, Victoria 3052, Australia}
\date{\today}
\maketitle
\begin{abstract}
We present some striking global consequences of a model quaternionic
quantum field theory which is locally complex.
We show how making the quaternionic structure a dynamical quantity
naturally leads to the prediction of cosmic strings and non-baryonic
hot dark matter candidates.
\end{abstract}
\pacs{\\{\it Report Number: UM--P--96/88; RCHEP 96/11.}}

\section{Introduction}
Of the various classes of phenomenological models and theoretical
constructs extending the standard model, one which has received
not much attention is quaternionic quantum mechanics.

This is curious as the theory proceeds from a firm axiomatic
basis \cite{bvn}, and previous researchers have made some profound
claims for its physical content including early models of unification
of fundamental forces \cite{fjss},
algebraic confinement of quarks or preons\cite{sla},
new effects in particle interferometry experiments \cite{peres,kgw},
and quaternionic effects in multi-particle correlated systems \cite{bja}.
Further, the theoretical construction is sufficiently restricted by
general mathematical principles that model building is constrained and
the theory can be predictive.

Important early work on the formulation
of quaternionic quantum mechanics and field theory was due to
Finkelstein, Jauch, Schiminovich and Speiser \cite{fjss},
who introduced the idea
of a gauge sector arising from a local, quaternionic structure in
space-time.  At the time of this proposal (the early 60's), the motivation
was to achieve a unification of electromagnetism with isospin.
With the success of the Salam--Weinberg electroweak unification, interest
in this work faded.  When divorced of its particle-physics interpretation,
however, the idea of introducing a locally complex description of nature
has several interesting consequences, which we shall develop in the next
section.

Our model is based on a theory which is locally equivalent to
the standard,
complex formalism, with the essentially quaternionic degrees of freedom
appearing as an additional, exotic gauge sector.  
Interaction with the Standard Model sector is constrained by the fact that
the new gauge fields are quaternion imaginary.  Hence, minimal coupling 
procedures fail, but these can be coupled to normal matter via 
non-renormalisable terms in an effective Lagrangian.  
While this can be criticised for being too na\"{\i}ve a construction, we
wish to bring our theory into the realm of experimentally constrainable
physics.

On this occasion, commemorating the 65$^{\rm th}$ birthday of
L.\ P.\ Horwitz, we feel it appropriate to add a personal note.
After the work of Finkelstein {\em et al.}, the next application of 
quaternionic quantum theory was proposed by Alder \cite{sla}, who 
raised the possibility of an algebraic mechanism for confinement in
hadronic physics.  Attention focussed on the properties of many-body
systems in a quaternionic framework.  A major investigation of 
the theory of tensor products in quaternionic quantum mechanical systems
was the classic paper of Horwitz and Biedenharn \cite{hb},
which analysed the
problems associated with forming tensor products of quaternion valued
single-particle wavefunctions.  This formalism, developed further in
\cite{hr},  lead the present authors to consider afresh the question
of uniquely quaternionic phenomena in quantum mechanical systems, the
result of which may be found in \cite{bja}, on a new effect in
multi-particle correlated systems.  The  impact of \cite{hb} and it's
successors on the quaternionic community has been profound, and it is
a privilege for us to contribute a paper to this volume.

\section{The model}

Studies of quaternionic quantum mechanics and field theory 
start with a  Hilbert space over the quaternion algebra.
The eigenvalues of operators need not commute, but
physical observables continue to be identified with Hermitian operators
which, as is well known, necessarily have real eigenvalues only.
(It is this fact which permits the existence of quaternionic quantum 
mechanics.)

The relation between kets and quaternion-valued wavefunctions 
follows the usual pattern (where the $\Phi_i$ are real),
\begin{equation}
\Phi(x_\mu ) \equiv \langle x_\mu |\Phi\rangle %
= \Phi_0 + \Phi_1 i + \Phi_2 j + \Phi_3 k \,.
\end{equation}

If we define $|\vec{\Phi}| = (\sum_{r=1}^{3}\Phi_{r}^{2})^{-1/2}$,
then there exists a pure imaginary quaternion of norm unity,
$\eta_{\Phi}$, such that
\begin{equation}
\Phi(x_{\mu}) = \Phi_{0}(x_{\mu})
+ |\vec{\Phi}(x_{\mu})|\eta_{\Phi}(x_{\mu})\,.
\end{equation}
From this we see that any two wavefunctions do not commute unless
their imaginary parts are parallel.  This property has confounded the
construction of a completely satisfactory tensor product, i.e., one which
would be quaternion linear in each factor, allow the definition of
a tensor product of operators on each factor, and admitting a positive
scalar product for the purpose of second quantisation of the theory
(see \cite{hr,nj} for comprehensive discussions). 

Alternatively, we can express $\Phi$ as an ordered pair of $i$-complex
numbers, in its so-called symplectic representation,
\begin{equation}
\Phi(x_\mu ) = \phi_{\alpha} + j \phi_{\beta} ,
\end{equation}
where $\phi_{\alpha} = \Phi_0 + i \Phi_1$ and
$\phi_{\beta} = \Phi_2 - i \Phi_3$.

Such a representation implicitly breaks the full quaternionic symmetry
of the theory. 
Given the requirement of eventually connecting with standard
(complex) phenomenology, there is a strong desire to introduce such a
breaking early in the theoretical formulation, often with a caveat 
that sufficiently early in the history of the universe the full 
quaternionic symmetry is to be restored.
Justification for introducing such a breaking follow from arguments
about the structure and symmetries of the equations of motion, from
the fact that a clever definition of the order of multiplication 
can make the second symplectic component (i.e., the $\phi_{\beta}$ term)
decay exponentially outside of interaction regions (making the theory
asymptotically $i$-complex), and from consideration of linearity
properties of tensor products.

We shall instead start with a model which, we maintain, retains
the spirit of the original while holding out a strong hope of connecting
with experimental physics.  We begin by summarising the locally complex
theory of Finkelstein {\em et al.}, but with our own interpretation, and
with an important addition.

Hence, consider a space-time on which there is defined a set of
scalar fields,
\begin{equation}
i(x_{\mu})\,,\quad j(x_{\mu})\,,\quad\mbox{and}\quad
k(x_{\mu}) \equiv i(x_{\mu})j(x_{\mu})\,,
\end{equation}
which locally define a basis for the set of pure imaginary quaternions,
\begin{equation}
i^2(x_{\mu}) = -1\,,\quad j^2(x_{\mu})=-1\,,\quad\mbox{and}\quad
\{i(x_{\mu}),j(x_{\mu})\}=0\,.
\end{equation}

We claim that the overwhelming success of conventional, complex
quantum mechanics and quantum field theory is strongly indicative of the
validity of a complex description of local physics (at least, on scales
presently accessible to laboratory physics).
Hence, in our model the matter and gauge fields will be assumed to be
$i(x)$-complex, i.e., amplitudes are real linear superpositions of 1 and
the $i(x_{\mu})$ field, and transform as vectors under a local
quaternionic gauge symmetry,
\begin{equation}
\phi(x_{\mu})\rightarrow q(x_{\mu})\phi(x_{\mu})q(x_{\mu})^{-1}\,,
\end{equation}
where $q$ is a pure imaginary quaternion of unit magnitude.

The gauge theory arising from this construction is SU(2) like.
We claim that this is a new sector;  we do not identify it with 
the fundamental SU(2) in electroweak theory, nor with any approximate
SU(2) in hadronic physics.   Instead, we follow the tradition of 
identifying essentially quaternionic physics with an outstanding puzzle
of contemporary physics, and claim that this new sector is a source 
of non-baryonic dark matter in the universe.  Further, we shall show
how the breaking of this local symmetry leads to cosmic strings.

The gauge-covariant derivative follows the usual pattern of Yang-Mills
theory, but with a quaternionic valued potential,
\begin{equation}
D_{\mu}\phi =\partial_{\mu}\phi +\frac{1}{2}(Q_{\mu}\phi -\phi Q_{\mu})\,,
\end{equation}
\begin{equation}
Q_{\mu}(x) \equiv \bar{A}_{\mu}(x) i(x) + B_{\mu}(x) j(x)
+ B'_{\mu}(x) k(x)\,.
\end{equation}
Our choice of notation follows, {\em a posteriori}, from our 
interpretation of the phenomenological implications of our model.

Finally, note that $D\phi$ reduces to $\partial\phi$ when $\phi$ is
real. 

Now we construct the field strength tensor,
\begin{equation}
K_{\mu\nu} = D_{\mu}Q_{\nu} - D_{\nu}Q_{\mu}\,,
\end{equation}
and exhibit the Lagrangian density for the new sector, which treats
$i$, $j$ and $k$ symmetrically,
\begin{equation}\label{Lfree}
{\cal L}_{\sc q} = \frac{1}{4}K^{\mu\nu}K_{\mu\nu}
+ \frac{\lambda}{2}|D_{\mu}i|^2 + \frac{\lambda'}{2}|D_{\mu}j|^2 \,.
\end{equation}

This Lagrangian differs from that of Finkelstein, {\em et
al.}\ \cite{fjss}
by the addition of the final term.  This requires several comments.

Firstly, it is
important to notice that the $i$ and $j$ fields are dimensionless.
Scale is introduced by the constants $\lambda$, $\lambda'$, which are
independent in the most general case, and have the dimension of 
mass squared.  In this regard, we might interpret the work of
Finkelstein {\em et al.}\/ as being in the $\lambda'\!\rightarrow 0$
limit.

As written, ${\cal L}_{\sc q}$ is composed of dimension 4 operators.
The components of the $Q$ potential have the canonical dimension of
vector bosons. We shall see that the presence of the dynamical 
terms $|Di|^2$ and $|Dj|^2$ inevitably give the bosons mass, and there
is no symmetric phase in which to renormalise the theory.  That is,
the magnitudes of $i$ and $j$ are fixed, and our theory is always
in the broken phase in which the massive vector bosons have (from
the UV behaviour of bosonic propagators) anomalous dimensions which 
renders the theory non-renormalisable \cite{iz}.
We do not regard this as catastrophic, our $i$ and $j$ fields have not 
been quantised, so our theory is of a semiclassical type.
A fully quantised theory of a dynamical quaternionic structure has
yet to be developed, and we concur with \cite{fjss} that any such theory
will share similarities with attempts to quantise gravity.  In fact,
interest in quantum gravity lead us to this field.

We have not included a $(\lambda''/2)|D_{\mu}k|^2$, as at each point
of space-time a consistent definition of the quaternionic algebra
requires $k(x) \equiv i(x)\,j(x)$, so there are no additional dynamical
degrees of freedom. 

In order to interpret the physical content of our model, we make a
particular local quaternionic gauge transformation, to what we call
the i gauge (strongly analogous to the unitary gauge of electroweak
physics).  Now there always exists a quaternionic gauge transformation
taking $i(x_{\mu})$ to a particular pure imaginary unit quaternion $i$,
$q_i(x_{\mu})\,i(x_{\mu})\,q_{i}(x_{\mu})^{-1} = i$ \cite{fjss}. 

In this particular gauge, the quaternionic potential is
\begin{equation}
Q_{\mu}(x) = \bar{A}_{\mu}(x)\,i + \left(B_{\mu}(x)
+ B'_{\mu}(x)\,i\right) j'(x)\,,
\end{equation}
where $j'(x)\!=\!q_i(x_{\mu})\,j(x)\,q_{i}(x_{\mu})^{-1}$
is defined up to a real phase,
\begin{equation}
j'(x) = \exp[\vartheta(x)\,i]\,j = \cos\vartheta(x)\,j + \sin\vartheta(x)\,k\,.
\end{equation}
That is, we have introduced a fixed basis $i$, $j$ and $k = ij$, 
and found that the remaining degrees of freedom reside in the 
phase of $j'$, which is a real Goldstone boson,
and in the transformed vector potentials
$\bar{A}$, $B$ and $B'$.  The Lagrangian has become
\begin{equation}
{\cal L}_{\sc q} = \frac{1}{4}K^{\mu\nu}K_{\mu\nu}
+ \frac{\lambda}{2}\left|\left[B_{\mu}j'
+ B'_{\mu}ij',\,i\right]\right|^2
+ \frac{\lambda'}{2}|D_{\mu}j|^2 \,.
\end{equation}
\[
= -\frac{1}{4}(\partial_{\mu}\bar{A}_{\nu} - \partial_{\nu}\bar{A}_{\mu}
+ B_{\mu}B'_{\nu} - B_{\nu}B'_{\mu})^2
+\{\mbox{permutations of $\bar{A}$, $B$ and $B'$}\}\:
\]
\begin{equation}
+ \frac{\lambda}{2}(B_{\mu}^2 + {B'_{\mu}}^{2})
+ \frac{\lambda'}{2}(\partial_{\mu}\vartheta + \bar{A}_{\mu})^2
+ \frac{\lambda'}{2}(B_{\mu}\sin\vartheta - B'_{\mu}cos\vartheta)^2
\,.\end{equation}

This particular gauge choice permits an interpretation of the physical
content of the theory.

First of all, we see that the dynamical nature of the $i$ field
has given degenerate masses to the $B$ and $B'$ fields,
in a type of spontaneous symmetry breaking noticed by \cite{fjss}.
If only the $i$ field
is regarded as being dynamical (the $\lambda'\rightarrow0$ limit),
then the SU(2) gauge symmetry is broken to an (Abelian) $\bar{A}$.
(A difference in the scales $\lambda >\lambda'$ will have
consequences for the phenomenology of our theory, and will be 
briefly discussed in the next section.)  Cosmological relic
populations of these $B$ and $B'$ fields will constitute 
warm dark matter candidates, as discussed in \cite{bhj}.  Also note that
the degeneracy of the masses is essentially due to the singling out
of $i$ as a special pure imaginary quaternion.

Note that symmetry breakdown has occurred via a choice of gauge,
which we maintain has to happen to remove the unphysical degrees of
freedom from the free field Lagrangian.  So a locally complex
quaternionic field theory has observational consequences which 
mimic those of an exotic SU(2) gauge theory in the usual 
complex framework, except that, on general grounds, coupling of
this sector to the standard sector is restricted to take a special
form described in the next section, and we find global phenomena
in the form of topological defects, as we now show.

The implications of the dynamical $j$ field are brought out most clearly
if we consider the Abelian restriction of our theory 
(the ``electromagnetic'' limit of \cite{fjss}), which is motivated
by requiring that $Di = 0$, in analogy with the consistency conditions
imposed on the Levi-Civita connection in classical general relativity
($\nabla g_{\mu\nu} = 0$, see \cite{rw}).  Then the $B$ and $B'$ fields
vanish identically, and we have the free field Lagrangian
\begin{equation}
{\cal L}_{{\sc q}\bar{\sc a}} 
= -\frac{1}{4}F_{\bar{\sc a}}^{\mu\nu}F_{\bar{\sc a}\mu\nu}
+ \frac{\lambda'}{2}(\partial_{\mu}\vartheta + \bar{A}_{\mu})^2\,,
\end{equation}
which has the pattern of the Abelian Higgs model \cite{no}, without
the usual polynomial potential term.  But our $j$ field is defined
to be of unit magnitude, and so we automatically have a vacuum 
expectation value for this field, which breaks the remaining
$\bar{A}$ Abelian symmetry. 

The consequences of this are that 
the $\bar{A}$ picks up a mass $\sqrt{\lambda'}$, and our theory
supports cosmic string solutions of the classic Nielsen--Olesen type.
That is, treating the $j$ field as a dynamical quantity on the same
footing as the $i$ field completes the breaking of the local,
quaternionic gauge symmetry, and introduces the possibility of 
global phenomena of a topological character. 
While cosmic strings might not be the most popular dark matter candidate
at the present date, it is interesting to see them arising in this 
context.

We contrast our treatment with the original of Finkelstein {\em et al.},
in which the electromagnetic limit removed all global effects.  Such a
construction can be criticised as being too close to the standard
complex theory, but suited well the motivations of the day. 

\section{Interaction with visible sector}
We now seek to couple this exotic sector to the locally complex fields
of the Standard Model.  We shall present our interaction  Lagrangian,
and discuss its physical content.  Detailed calculations using this
model are reported elsewhere \cite{bhj}. 

We follow the general discussion of \cite{hol}, which  discusses possible
interaction Lagrangian terms in the context of axionic physics.
In our case, $Q$ and $K$ are pure imaginary
quaternion, which leads us to propose an effective interaction Lagrangian 
which couples field densities to $K^2$,
\begin{equation}
{\cal L}_{\rm int} =
\frac{g_{\sc q}}{4\Lambda_{\sc q}^2}\phi^{\dag}\phi\,K^2
+ \frac{g_{\sc q}}{4\Lambda_{\sc q}^3}\bar{\psi}\psi\,K^2
+ \frac{g_{\sc q}}{16\Lambda_{\sc q}^4}\,{\rm tr}F^2\,K^2\,.
\end{equation}
The exclusion
of terms such as $\bar{\psi}\sigma\psi\cdot K$ is a constraint on model
building imposed by the framework of our theory.

Here, $\lambda$ and $\lambda'$ are {\em a priori}\/ different.
and we propose that $\lambda' \ll \lambda$.
Then in a low to middle energy system, we might expect $\bar{A}$
production to dominate.

The scale $\Lambda_{\sc q}$ is a characteristic quantity in QQFT,
with a role that is equivalent to that of $\Lambda_{\sc qcd}$
in the $\overline{{\rm MS}}$ renormalisation schemes.  Our theory is only
effective.  Consideration of the underlying fundamental theory is
of paramount interest, but is expected to give equivalent results
to the effective theory below the $\Lambda_{\sc q}$ scale.  It is
natural to expect $\Lambda_{\sc q}$ to be somewhere between the 
electroweak and Planck scales, but we do not rule out the possibility
that this effective description may be inadequate for a description 
of top quark physics.  This would place nontrivial predictions
of our theory 
in an experimentally interesting region.  These interaction terms 
are contact terms, describing mutli-particle production above a mass
pole (governed by $\lambda'$) with $\Lambda_{\sc q}^{-1}$ our perturbation
parameter (more precisely, some relevant mass $m \ll \Lambda_{\sc q}$
allows us to take $m/\Lambda_{\sc q}$ as our parameter).

Note that several generalisations of the Standard Model immediately
suggest extensions of the above scheme.  For example, similar couplings
of SUSY partners to SM fields, or models with horizontal symmetry.  We
leave development of these ideas for the future.

Our interpretation of the physical content of this model is consistent
with the introduction of non-baryonic dark matter.  Coupling of normal
matter to the new fields is suppressed by the $\Lambda_{\sc q}$ scale,
and it is possible to investigate the implications of this dark matter
in astrophysical systems and at cosmological scales.  Elsewhere, we
present a detailed discussion of such an investigation.

\section{Summary and conclusion}
We have introduced a local quaternionic gauge structure onto space-time.
Treating $i$ and $j$ as dynamical fields breaks this symmetry completely,
leaving three massive vector bosons.  Cosmic string solutions exist,
arising from the orientation of $j$.

Our theory is non-renormalisable.  It is a theory of vector bosons
and dimensionless scalar fields, which recalls semi-classical treatments
of gravity.  After transforming to the i gauge, we find that the 
quaternionic symmetry takes the form of an exotic SU(2) gauge theory
in the standard complex framework, with global phenomena appearing
in the form of cosmic strings.

Coupling this quaternionic sector to the Standard Model sector has
only been achieved at the level of an effective theory, which is 
constrained by the quaternionic origin of the exotic bosons to be
of a nonrenormalisable form.

\end{document}